\documentclass[pre,twocolumn,aps,showpacs,amsmath,amssymb]{revtex4-1}
\usepackage{color}
\usepackage{bm}
\usepackage{graphicx}
\usepackage{braket}

\usepackage[plainpages=false,pdfpagelabels,colorlinks=true,linkcolor=red,urlcolor=blue,citecolor=blue,pdftitle={},pdfauthor={},pdfdisplaydoctitle=true,pdfduplex=DuplexFlipLongEdge]{hyperref} 

\def\Tr{\mbox{Tr}\,}

\begin{document}
\title{Work extraction in an isolated quantum lattice system: \\ Grand canonical and generalized Gibbs ensemble predictions}
\author{Ranjan Modak} 
\author{Marcos Rigol}
\affiliation{Department of Physics, Pennsylvania State University, University Park, Pennsylvania 16802, USA}

\begin{abstract}
We study work extraction (defined as the difference between the initial and the final energy) in noninteracting and (effectively) weakly interacting isolated fermionic quantum lattice systems in one dimension, which undergo a sequence of quenches and equilibration. The systems are divided in two parts, which we identify as the subsystem of interest and the bath. We extract work by quenching the on-site potentials in the subsystem, letting the entire system equilibrate, and returning to the initial parameters in the subsystem using a quasi-static process (the bath is never acted upon). We select initial states that are direct products of thermal states of the subsystem and the bath, and consider equilibration to the generalized Gibbs ensemble (GGE, noninteracting case) and to the Gibbs ensemble (GE, weakly interacting case). We identify the class of quenches that, in the thermodynamic limit, results in GE and GGE entropies after the quench that are identical to the one in the initial state (quenches that do not produce entropy). Those quenches guarantee maximal work extraction when thermalization occurs. We show that the same remains true in the presence of integrable dynamics that results in equilibration to the GGE.
\end{abstract}
\maketitle

\section{Introduction} 

Work is a familiar concept in the context of classical thermodynamics, which deals with systems with a large number of particles. A goal of classical thermodynamics is to identify protocols that provide an efficient way of converting heat into work. In recent years, there has been a lot of interest in developing a thermodynamic framework to deal with small systems that can be far from equilibrium~\cite{jarzynski2011equalities, seifert2012stochastic}. In particular, following seminal work by Jarzynski~\cite{jarzynski1997nonequilibrium}, fluctuation theorems have become an area of intense activity. Fluctuation theorems have been generalized to understand the stochastic fluctuations of work done on non-equilibrium quantum systems~\cite{crooks1999entropy, kurchan2000quantum, tasaki2000jarzynski, campisi2011colloquium}. Another recent development comes from the so-called single-shot information theory, which was initially postulated to study finite-size effects in quantum cryptography. Single-shot information theory has become a useful tool in understanding work extraction in the context of quantum thermodynamics~\cite{aaberg2013truly, horodecki2013fundamental}. 

In parallel, extraordinary advances in experiments with ultracold atomic gases~\cite{cazalilla_11, bloch_dalibard_08} have motivated much research on the far-from-equilibrium dynamics and the description after equilibration of isolated many-body quantum systems \cite{d2016quantum, eisert_friesdorf_15, polkovnikov_sengupta_11}. Of particular interest has been the dynamics following the so-called quantum quenches \cite{calabrese_cardy_06}, in which the system is initially in a stationary state of some time-independent Hamiltonian and that Hamiltonian is suddenly changed into a new one that is also time-independent. If the Hamiltonian after the quench is quantum chaotic, i.e., if its distribution of many-body energy level spacings is of the Wigner-Dyson type, one expects thermalization to occur~\cite{santos2010onset, d2016quantum}. Namely, one expects that after equilibration observables are described by traditional statistical mechanics \cite{d2016quantum, rigol2008thermalization, rigol_14}. This can be understood to be a consequence of eigenstate thermalization \cite{deutsch1991quantum, srednicki1994chaos, rigol2008thermalization, rigol2012alternatives}. On the other hand, if the Hamiltonian after the quench is integrable, one expects generalized thermalization to occur. Namely, one expects that after equilibration observables are described by a generalized Gibbs ensemble (GGE), which takes into account the presence of an extensive set of nontrivial conserved quantities \cite{rigol2007relaxation, calabrese_essler_11, ilievski_denardis_15} (see Refs.~\cite{vidmar_rigol_16, essler_fagotti_16, cazalilla_chung_16, caux_16} for recent reviews). This can be understood to be a consequence of generalized eigenstate thermalization \cite{cassidy2011generalized, vidmar_rigol_16, caux_essler_13}.

Here, we explore work extraction in isolated (integrable) noninteracting and (effectively) weakly interacting fermionic quantum lattice systems in one dimension as described by a quadratic Hamiltonian. The isolated systems are divided in two parts, which we identify as the subsystem of interest and the bath. The specific protocol we consider is motivated by a possible straightforward implementation in experiments, and consists of: (i) quenches of on-site potentials in the subsystem; (ii) equilibration to the GGE (noninteracting case) or the grand canonical ensemble (GE, weakly interacting case); and (iii) return to the initial values of the on-site potentials in the subsystem by means of quasi-static process with equilibration to the GGE and the GE (the bath is never acted upon). The work extracted is computed as the difference between the initial and the final energy of the entire system. Since the average energy of a thermal state (with non-negative temperature) can only increase due to unitary operations (quantum quenches), because of passivity~\cite{pusz1978passive, lenard1978thermodynamical}, the initial states cannot be thermal equilibrium states of the entire system (we would like to be able to extract work). Instead, they are selected to be direct products of grand canonical states of the subsystem and the bath at the same temperature but at different chemical potentials (different site occupations). 

In a recent study, Perarnau-Llobet {\it et al.}~\cite{perarnau2016work} discussed upper bounds for the work that can be extracted in processes involving equilibration to the GE or the GGE in isolated quantum systems. In the context of our protocol, we identify a class of quenches that do not produce entropy when equilibration occurs to the GE or the GGE, which automatically ensures maximal work extraction for equilibration to the GE. We show that those quenches saturate the bound for work extraction under equilibration to the GGE. 

The paper is organized as follows. In Sec.~\ref{secII}, we introduce the model, quench protocol, and the initial states considered. We discuss the results for equilibration to the GE in Sec.~\ref{secIII}, and for equilibration to the GGE in Sec.~\ref{secIV}. In Sec.~\ref{secV}, we present a comparison between the results obtained for the GE and for the GGE. A summary of our results is presented in Sec.~\ref{secVI}. 

\section{Model, quench protocol, and initial states\label{secII}}

We study noninteracting (and, effectively, weakly interacting) fermions in one-dimensional (1D) lattices with open boundary conditions. The system is divided in two parts, which we identify as the subsystem of interest and the bath. They are described by the tight-binding quadratic Hamiltonians $\hat{\mathcal H}_{s}$ and $\hat{\mathcal H}_{b}$, respectively
\begin{eqnarray}
 \hat{\mathcal H}_{s}(V_s)&=&-\sum _{i=1}^{L_{s}-1}(\hat{c}^{\dag}_i\hat{c}^{}_{i+1}+\text{H.c.})+V_{s}\sum _{i=1}^{L_{s}}\hat{n}_i,  \\
 \hat{\mathcal H}_{b}(V_b)&=&-\sum _{i=L_{s}+1}^{L-1}(\hat{c}^{\dag}_i\hat{c}^{}_{i+1}+\text{H.c.})+V_{b}\sum _{i=L_{s}+1}^{L}\hat{n}_i, \nonumber
\end{eqnarray}
where $\hat{c}^{\dag}_i$ ($\hat{c}_{i}$) is the fermionic creation (annihilation) operator at site $i$, $\hat{n}_i = \hat{c}^{\dag}_i\hat{c}_{i}$ is the number operator, and $L_s$ ($L$) is the size of the subsystem (entire system). We have set the hopping amplitudes in the subsystem and bath to unity, and $V_{s}$ ($V_{b}$) is the on-site potential of the subsystem (bath). Dynamics are studied under the total Hamiltonian
\begin{equation}\label{eq:finHam}
 \hat{\mathcal H}(V_s,V_b)=\hat{\mathcal H}_s(V_s)+\hat{\mathcal H}_b(V_b)-(\hat{c}^{\dag}_{L_{s}}\hat{c}^{}_{L_{s}+1}+\text{H.c.}).
\end{equation}

\begin{figure}
\centering
\includegraphics[width=0.48\textwidth]{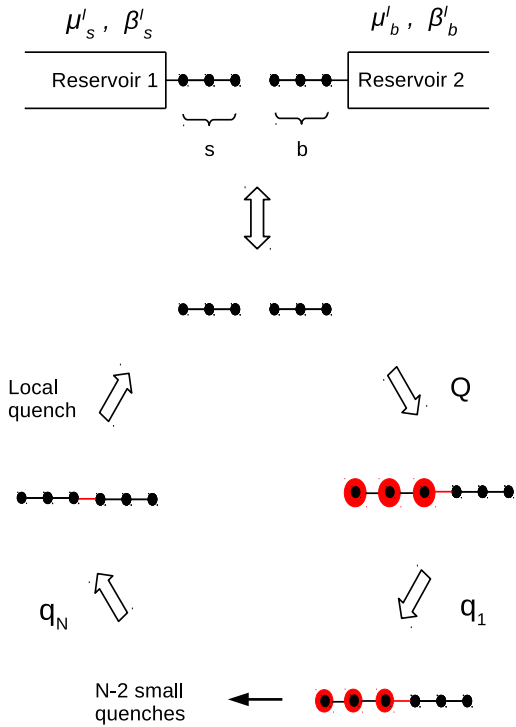}
\vspace{-0.5 in}
\caption{(Color online) Sketch of the cyclic process studied in this work (see text for the description).}
\label{fig0}
\end{figure}

We prepare the initial state to be a direct product of GE density matrices of the subsystem and bath with $\hat{\mathcal H}_{s}(V^I_s)$ and $\hat{\mathcal H}_{b}(V_b)$, respectively (see Sec.~\ref{secIIa}). [This can be done by (weakly) connecting the subsystem and the bath to two reservoirs (see Fig.~\ref{fig0}).] Next, we connect the subsystem and the bath, and at the same time quench the on-site potentials of the subsystem from $V^{I}_{s}$ to $V^F_{s}$ (see Fig.~\ref{fig0}). The entire system is then allowed to equilibrate to the GE and the GGE under $\hat{\mathcal H}(V^F_s,V_b)$. After equilibration, which is ensured by taking the density matrix of the entire system to be the appropriate GE or GGE density matrix, we apply a large number $N$ of weak quenches followed by equilibration to the GE or the GGE. In each of those $N$ weak quenches, the on-site potential in the subsystem is changed by $(V^{I}_{s}-V^{F}_{s})/N$, so that at the end we have the subsystem at the initial value $V^{I}_{s}$. In a final ($N+1$) quench, we turn off the hopping between the subsystem and the bath and let the system equilibrate to the GE or the GGE. Note that the latter is a local quench, i.e., it does not produce extensive changes in thermodynamic quantities. This completes our cyclic process (see Fig.~\ref{fig0}). (One can prepare again the initial direct product of GE density matrices by connecting the subsystem and the bath to the two reservoirs.)

Relaxation to the GGE is assumed in order to describe what happens under the dynamics dictated by the quadratic Hamiltonian $\hat{\mathcal H}(V_s,V_b)$. It is actually straightforward to prove that the infinite-time average of the entire one-body density matrix of a noninteracting system (from which all observables in our noninteracting system can be computed) is, in the absence of degeneracies in the single-particle spectrum, identical to that of the GGE \cite{he2013single, ziraldo_santoro_13}. However, the density matrix of the entire system, at any time after a quench, is never that of the GGE (the dynamics is unitary). This is also true for the one-body density matrix, because the fermions are noninteracting \cite{wright_rigol_14}. Hence, one may be wary about replacing the density matrix ``after equilibration'' by that of the GGE. We report numerical results that support the appropriateness of this procedure (see also Ref.~\cite{perarnau2016work}), as the exact time evolution in which one waits random times after equilibration and the GGE replacement produce nearly identical results for the work extraction.

Relaxation to the GE is assumed in order to describe what happens under the dynamics dictated by the quadratic Hamiltonian $\hat{\mathcal H}(V_s,V_b)$ plus very weak integrability-breaking interactions. We imagine the weak integrability-breaking interactions as allowing the systems to thermalize at long times \cite{rigol_14, bertini_essler_15, brandino_caux_15, rigol2016fundamental, bertini_essler_16}, while being weak enough not to significantly change the thermal expectation value of macroscopic observables (such as the energy, which is needed to compute the work extracted) from the result in the noninteracting limit. The fact that one can replace the density matrix of the time-evolving state of a quantum chaotic system after equilibration, without needing to wait random times as in the integrable case, by that of the GE has been discussed in Refs.~\cite{ji_fine_2011, ikeda_sukumichi_15, d2016quantum}.

The work extracted due to our cyclic process, $W$, is defined as
\begin{equation}\label{eq:work}
 W=\Tr\left[(\hat{\rho}^I-\hat{\rho}^F)\,[\hat{\mathcal H}_s(V^I_s)+\hat{\mathcal H}_b(V_b)]\right],
\end{equation}
which is the difference between the energy in the initial and final states~\cite{alicki2013entanglement, d2016quantum}, where $\hat{\rho}^I$ ($\hat{\rho}^F$) is the density matrix of the initial (final) state. For all calculations reported here, we take $V^{I}_{s}=V_{b}=0$ and $V^{F}_s=V>0$.

\subsection{Initial States\label{secIIa}}

We consider initial states that are product states of the subsystem and the bath, i.e., whose density matrix can be written as
\begin{equation}
 \hat{\rho}^I = \hat{\rho}^I_s \otimes \hat{\rho}^I_b.
\end{equation}
We take the density matrices of the subsystem and the bath to be grand canonical, 
\begin{eqnarray}
 \hat{\rho}^I_s&=&\exp(-\beta^I_s [\hat{\mathcal H}_s(0)-\mu^I_s \hat{\mathcal N}_s])/Z^I_s,\nonumber\\
 \hat{\rho}^I_b&=&\exp(-\beta^I_b [\hat{\mathcal H}_b(0)-\mu^I_b \hat{\mathcal N}_b])/Z^I_b,
\end{eqnarray}
respectively. $\hat{\mathcal N}_s$ ($\hat{\mathcal N}_b$) is the total number of particles operator of the subsystem (bath), $\beta^I_s$ ($\beta^I_b$) and $\mu^I_s$ ($\mu^I_b$) are the inverse temperature and chemical potential of the subsystem (bath), respectively, and $Z^I_s$ ($Z^I_b$) is the grand-canonical partition of the subsystem (bath)~\cite{he2012initial}:
\begin{equation}
Z^I_{s(b)}=\prod _{i=1}^{L_{s(b)}}\left[1+e^{-\beta^I_{s(b)}\big(\epsilon_{s(b)}^i-\mu^I_{s(b)}\big)}\right],
\end{equation}
where $\epsilon_{s(b)}^i$ are the single-particle eigenenergies of the subsystem (bath). The total energy and particle number of the subsystem (bath) are
\begin{equation}
 E^I_{s(b)}=\Tr[\hat{\rho}^I_{s(b)}\hat{\mathcal H}^I_{s(b)}]\ \ \text{and}\ \ 
 {\mathcal N}^I_{s(b)}=\Tr[\hat{\rho}^I_{s(b)}\hat{\mathcal N}_{s(b)}], \label{eq:EN}
\end{equation}
respectively. The occupation per site in the subsystem is then $n^I_s={\mathcal N}^I_s/L_s$.

If the chemical potential and the inverse temperature of the subsystem and the bath were chosen to be same, then $\hat{\rho}^I$ is the thermal equilibrium state of 
\begin{equation}\label{eq:HI}
 \hat{\mathcal H}^I\equiv\hat{\mathcal H}_s(0)+\hat{\mathcal H}_b(0),
\end{equation}
which is the initial and final Hamiltonian of our cyclic process. Since thermal equilibrium states are passive, one would not be able to extract work in a cyclic process starting from such an initial state~\cite{d2016quantum}. Hence, one needs to choose $\mu^I_s \neq \mu^I_b$ and/or $\beta^I_s \neq \beta^I_b$. Next we show that, in order to be able to extract work with our quench protocol, we need $\mu^I_s <\mu^I_b$.

Since the initial state is a product of two thermal states, its entropy can be written as
\begin{equation}
 S^I= S^I_{s}+S^I_b,\label{eq:entI}
\end{equation}
where $S^I_{s(b)}$ is the GE entropy of the subsystem (bath)~\cite{he2012initial}: 
\begin{equation}
 S^I_{s(b)}=\ln Z^I_{s(b)}+\beta^I_{s(b)} [E^I_{s(b)}-\mu^I_{s(b)} {\mathcal N}^I_{s(b)}].
\end{equation}
Given the total initial energy and the total initial number of particles
\begin{equation}\label{eq:ENI}
 E^I= E^I_{s}+E^I_{b}\quad\text{and}\quad {\mathcal N}^I\equiv {\mathcal N}^I_{s}+{\mathcal N}^I_{b},
\end{equation}
respectively, one can construct the density matrix $\hat{\rho}^\text{GE}_I$ of a thermal state of the entire system that matches the initial total energy and number of particles. Namely, $\hat{\rho}^\text{GE}_I=\exp[-\beta_I' (\hat{\mathcal H}^I-\mu_I' \hat{\mathcal N})]/\Tr[\exp(-\beta_I' [\hat{\mathcal H}^I-\mu_I'\hat{\mathcal N}])]$, such that $E^I=\Tr[\hat{\rho}^\text{GE}_I \hat{\mathcal H}^I]$ and ${\mathcal N}^I=\Tr[\hat{\rho}^\text{GE}_I \hat{\mathcal N}]$. Since $V^{I}_{s}=V_{b}=0$, that state has a uniform occupation of the sites, $n_I={\mathcal N}^I/L$. We call the entropy corresponding to such a thermal state $S^\text{GE}_I$.

Now we can consider the strong quench of the on-site potential ($V_s^I=0\rightarrow V_s^F=V$) starting from $\hat{\rho}^I$ and $\hat{\rho}^\text{GE}_I$. They result in energies after the quench
\begin{equation}\label{eq:EQ}
 E_{Q}=E^I+V n^I_s L_s\quad\text{and}\quad E'_{Q}=E^I+V n_I L_s,
\end{equation}
respectively. ($n^I_s$ and $n_I$ are the site occupations in the subsystem for $\hat{\rho}^I$ and $\hat{\rho}^\text{GE}_I$, respectively.) For $V>0$ and $n^I_s \geq n_I$,  $E_{Q} \geq E'_{Q}$. Since the entropy in the GE is a monotonic function of the energy, we immediately realize that equilibration to the GE results in entropies $S^\text{GE}_{Q}$ and $S^{'\text{GE}}_{Q}$, corresponding to $E_{Q}$ and $E'_{Q}$, respectively, which satisfy $S^\text{GE}_{Q} \geq S^{'\text{GE}}_{Q}$. We also know that, as a result of the quench, $S^{'\text{GE}}_{Q} \geq S^\text{GE}_I$, from which it follows that $S^\text{GE}_{Q}\geq S^\text{GE}_I$.

Finally, let us consider the quasi-static process that brings the system back to the initial Hamiltonian. In the limit $N\rightarrow\infty$, when equilibration to the grand-canonical ensemble is assumed in every weak quench, the entropy of the thermal state at the end of the quasi-static process $S^\text{GE}_F$ is $S^\text{GE}_F\simeq S^\text{GE}_{Q}$~\cite{aizenman1981third}. Additionally, we see that, for $V>0$ and $n^I_s \geq n_I$, $S^\text{GE}_F\geq S^\text{GE}_I$. Again, since the entropy in the GE is a monotonic function of the energy, we conclude that the final energy of the system after the cyclic process is larger than the initial energy. As a result, no work can be extracted [$W$ in Eq.~\eqref{eq:work} is negative]. Hence, in order to be able to extract work, we need $n^I_s < n_I$. This can be ensured by choosing $\mu^I_s <\mu^I_b$. Note that our analysis and conclusion are completely independent of the values of $\beta^I_s$ and $\beta^I_b$. In what follows, we take $\beta^I_s=\beta^I_b=\beta^I$ for all our calculations. This choice plays an important role in devising the protocol that maximizes the work extracted. 

\section{Grand Canonical Ensemble (GE)\label{secIII}}

For isolated integrable quantum systems, such as those described by quadratic Hamiltonians, the expectation values of observables after equilibration following a quench are not described by traditional ensembles of statistical mechanics. (This is true even if the initial state before the quench is a thermal equilibrium state of a quantum chaotic Hamiltonian \cite{rigol2016fundamental}.) The reason for this lack of thermalization is the presence of an extensive set of nontrivial conserved quantities. However, very weak integrability-breaking interactions are expected to ensure that the system thermalizes at long times \cite{rigol_14, bertini_essler_15, brandino_caux_15, rigol2016fundamental, bertini_essler_16}, even if they do not significantly change the energy of the system from their noninteracting values. 

With this in mind (see also Sec.~\ref{secII}), in this section we replace the density matrix of the entire system after a quench with that of a GE whose energy and total number of particles match those of the noninteracting system after the quench. That GE density matrix is assumed to describe observables of interest here, such as site occupations, after equilibration in the presence of weak integrability-breaking interactions. What happens in the absence of interactions is the subject of the next section. Studying quadratic Hamiltonians allows us to gain analytic insights about the specific quench protocol that saturates theoretical bounds in the thermodynamic limit. It also allows us to study numerically finite systems that can be small or as large as desired.

\subsection{Quasi-static process in the thermodynamic limit\label{secIIIa}}

Given our definition of work extracted [see Eq.~\eqref{eq:work}], maximal work is associated with minimal energy after completing the cyclic process. Assuming the system thermalizes, this corresponds to the case in which the system has minimal GE entropy at the end of the cyclic process. Therefore, our goal in order to maximize work is to design a cyclic process that keeps the entropy constant after every quench (the entropy cannot decrease). If this is achieved, we saturate the upper bound for the work that can be extracted in a cyclic process~\cite{perarnau2016work}
\begin{equation}\label{eq:maxwork}
 W^\text{GE}_\text{max}=\Tr\left[(\hat{\rho}^I -\hat{\rho}^\text{GE}_\text{max})\,[\hat{\mathcal H}_s(0)+\hat{\mathcal H}_b(0)]\right],
\end{equation}
where $\hat{\rho}^\text{GE}_\text{max}$ is the density matrix of a GE that has the same entropy as the initial state, $S^\text{GE}_\text{max}=S^I$. (It also must have the same number of particles, but this is enforced no matter the protocol implemented.) In the limit $N\to \infty$ (quasi-static protocol), and for large systems (we note that equilibration times increase with increasing system size),  the entropy at the end of the cyclic process equals that after the strong quench, i.e., $S^\text{GE}_F \simeq S^\text{GE}_{Q}$. Hence, all we need to do is to find a protocol by means of which the GE entropy after the strong quench is that of the initial state, $S^\text{GE}_{Q}=S^I$.

In the thermodynamic limit, when $L_s\to \infty$ and $L_b\to \infty$ for $\eta=L_s/L$ finite, the initial site occupations of the subsystem and the bath can be obtained as
\begin{eqnarray}
 n^I_s&=&\int_{-2t}^{2t}g(\epsilon)\frac{1}{\exp[\beta^I(\epsilon -\mu^I_s)]+1} d\epsilon ,\nonumber \\
 n^I_b&=&\int_{-2t}^{2t}g(\epsilon)\frac{1}{\exp[\beta^I(\epsilon -\mu^I_b)]+1} d\epsilon , 
 \label{Eq.3}
\end{eqnarray}
where $g(\epsilon)=(\pi \sqrt {4t^2 -\epsilon ^{2}})^{-1}$ is the density of states. Similarly, the initial energies of the subsystem and the bath read
\begin{eqnarray}
 E^I_s&=&L_s\int_{-2t}^{2t}\epsilon\, g(\epsilon)\frac{1}{\exp[\beta^I(\epsilon -\mu^I_s)]+1} d\epsilon, \nonumber \\
 E^I_b&=&L_b\int_{-2t}^{2t}\epsilon\, g(\epsilon)\frac{1}{\exp[\beta^I(\epsilon -\mu^I_b)]+1} d\epsilon.
 \label{Eq.4}
\end{eqnarray}

After the quench $V_s^I=0\rightarrow V_s^F=V$, the GE density matrix of the entire system is 
\begin{equation}
 \hat\rho_Q^\text{GE}= \exp(-\beta_Q [\hat{\mathcal H}(V,0)-\mu_Q \hat{\mathcal N}])/Z_Q,
\end{equation}
where $Z_Q=\Tr[\exp(-\beta_Q [\hat{\mathcal H}(V,0)-\mu_Q \hat{\mathcal N}])]$. $\beta_Q$ and $\mu_Q$ are computed such that the GE energy and number of particles match the results after the quench, namely, $E_{Q}$ [see Eq.~\eqref{eq:EQ}] and ${\mathcal N}^I$ [see Eq.~\eqref{eq:ENI}], respectively.

Within the local density approximation, the GE energy of the subsystem and the bath after the quench can be obtained as
\begin{eqnarray}
 E^\text{GE}_s&=&L_s\int_{-2t+V}^{2t+V}\epsilon\, g(\epsilon-V)\frac{1}{\exp[\beta_{Q}(\epsilon-\mu_{Q})]+1}  d\epsilon,\nonumber\\
 E^\text{GE}_b&=&L_b\int_{-2t}^{2t}\epsilon\, g(\epsilon)\frac{1}{\exp[\beta_{Q}(\epsilon-\mu_{Q})]+1}d\epsilon,
\end{eqnarray}
respectively. We can rewrite the energy of the subsystem as
\begin{eqnarray}
 E^\text{GE}_s&=&L_s\int_{-2t}^{2t}\epsilon g(\epsilon)\frac{1}{\exp[\beta_{Q}(\epsilon -\mu_{Q}+V)]+1}  d\epsilon  \label{Eq.5} \\
 &&+VL_s\int_{-2t}^{2t} g(\epsilon)\frac{1}{\exp[\beta_{Q}(\epsilon -\mu_{Q}+V)]+1} d\epsilon \nonumber,
\end{eqnarray}

Neglecting the $O(1)$ contribution of the hopping between the subsystem and the bath, the total energy
\begin{equation}\label{matchE}
 E^\text{GE}_s+E^\text{GE}_b=E^I_s+E^I_b+V n^I_s L_s
\end{equation}
[see Eqs.~\eqref{eq:ENI} and~\eqref{eq:EQ}], where $E^I_s$ and $E^I_b$ can be computed using Eq.~\eqref{Eq.4}, and $n^I_s$ can be computed using Eq.~\eqref{Eq.3}. A trivial solution to Eq.~\eqref{matchE} is obtained for $E^\text{GE}_s=E^I_s+V n^I_s L_s$ and $E^\text{GE}_b=E^I_b$, which require $\beta_{Q}=\beta^I$, $\mu_{Q}=\mu^I_b$, as well as $V=\mu^I_b-\mu^I_s$. 

This solution trivially satisfies that the total number of fermions after the quench remains the same as before the quench, and that the entropy of the initial state
\begin{eqnarray}
 S^I&=&L_s\int_{-2t}^{2t} g(\epsilon)\,\Pi\left(\frac{1}{\exp[\beta^I(\epsilon -\mu^I_s)]+1}\right) d\epsilon \label{entropy_lda0}\\
 &&+L_b\int_{-2t}^{2t} g(\epsilon)\,\Pi\left(\frac{1}{\exp[\beta^I(\epsilon -\mu^I_b)]+1}\right) d\epsilon, \nonumber
\end{eqnarray}
and of the GE describing the system after the quench
\begin{eqnarray}
 S^\text{GE}_{Q}&=&L_s\int_{-2t}^{2t} g(\epsilon)\,\Pi\left(\frac{1}{\exp[\beta_{Q}(\epsilon - \mu_{Q}+V)]+1}\right) d\epsilon \nonumber \\
 &+&L_b\int_{-2t}^{2t} g(\epsilon)\,\Pi\left(\frac{1}{\exp[\beta_{Q}(\epsilon -\mu_{Q})]+1}\right) d\epsilon,
 \label{entropy_lda}
\end{eqnarray}
where $\Pi(x)=-x\ln x-(1-x)\ln (1-x)$, are the same. This is possible because the initial state of our quench, which is not a thermal equilibrium state of the initial Hamiltonian, is very close to a thermal equilibrium state of the Hamiltonian after the quench. Such quenches can be implemented in a wide range of settings, including interacting systems.

A straightforward example in the context of the quadratic Hamiltonian \eqref{eq:HI} is the case in which initially the subsystem and the bath have the same chemical potential $\mu^I_b=\mu^I_s=0$ but different inverse temperature ($\beta^{I}_{s}$ and $\beta^{I}_b$, respectively). In this case, one can extract work by quenching the hopping amplitudes in the subsystem (no quench of the on-site potentials). Maximal work can be extracted for a strong quench of the hopping amplitude in the subsystem from $t\rightarrow t_Q$ with $t_Q/t=\beta^{I}_{s}/\beta^{I}_b$, and then returning to $t$ using a quasi-static process ($N$ weak quenches in the subsystem).

\subsection{Work extraction and entropy differences vs the number of quenches}

Next, we would like to gain an understanding of what happens in finite systems and for a finite number of quenches. For this, we use numerical calculations. Since the Hamiltonian of interest here is quadratic, all observables in thermal equilibrium can be computed from the one-body density matrix, $\rho^\text{GE}_{ij}=\Tr[\hat{\rho}^\text{GE} \hat{c}^{\dag}_i \hat{c}^{}_{j}]$, which can be obtained as
\begin{equation}
 \rho^\text{GE}_{ij}=\delta_{ij}-\left[I+e^{-\beta(H-\mu)}\right]_{ji}^{-1}\, ,
\label{eq:OBDM_FT_ii}
\end{equation}
where $I$ is the identity matrix, and $H$ is the matrix representing our Hamiltonian in the single-particle basis $\hat{\mathcal H}=\sum_{ij} \hat{c}^{\dag}_i H_{ij}\hat{c}^{}_{j}$ \cite{muramatsu_99, assaad_02, rigol2005finite}. 

Given the total energy $E_{Q}$ [see Eq.~\eqref{eq:EQ}], and the GE site occupancy in the subsystem $n^\text{GE}_{s,Q}$, after the quench $V_s^I=0\rightarrow V_s^F=V$, it is straightforward to see that the energy of the entire system after the $N$ weak quenches in which thermalization occurs, $E^\text{GE}_{F,N}$, depends on $N$ [in this analysis we ignore the final ($N+1$) local quench in which the subsystem and the bath are disconnected]. Denoting the site occupation in the subsystem after thermalization following the $m^\text{th}$ weak quench as $n^\text{GE}_{s,q_m}$, one can write
\begin{eqnarray}
 E^\text{GE}_{F,1}&=&E_{Q}-VL_sn^\text{GE}_{s,Q} \nonumber \\
 E^\text{GE}_{F,2}&=&E_{Q}-\frac{V}{2}L_s(n^\text{GE}_{s,Q}+n^\text{GE}_{s,q_1}) \nonumber \\
 &&\ldots \\
 &&\ldots \nonumber \\
 E^\text{GE}_{F,N}&=&E_{Q}-\frac{V}{N}L_s(n^\text{GE}_{s,Q}+n^\text{GE}_{s,q_1}+...+n^\text{GE}_{s,q_{N-1}}). \nonumber
\end{eqnarray}

\begin{figure}[!t]
\includegraphics[width=0.48\textwidth]{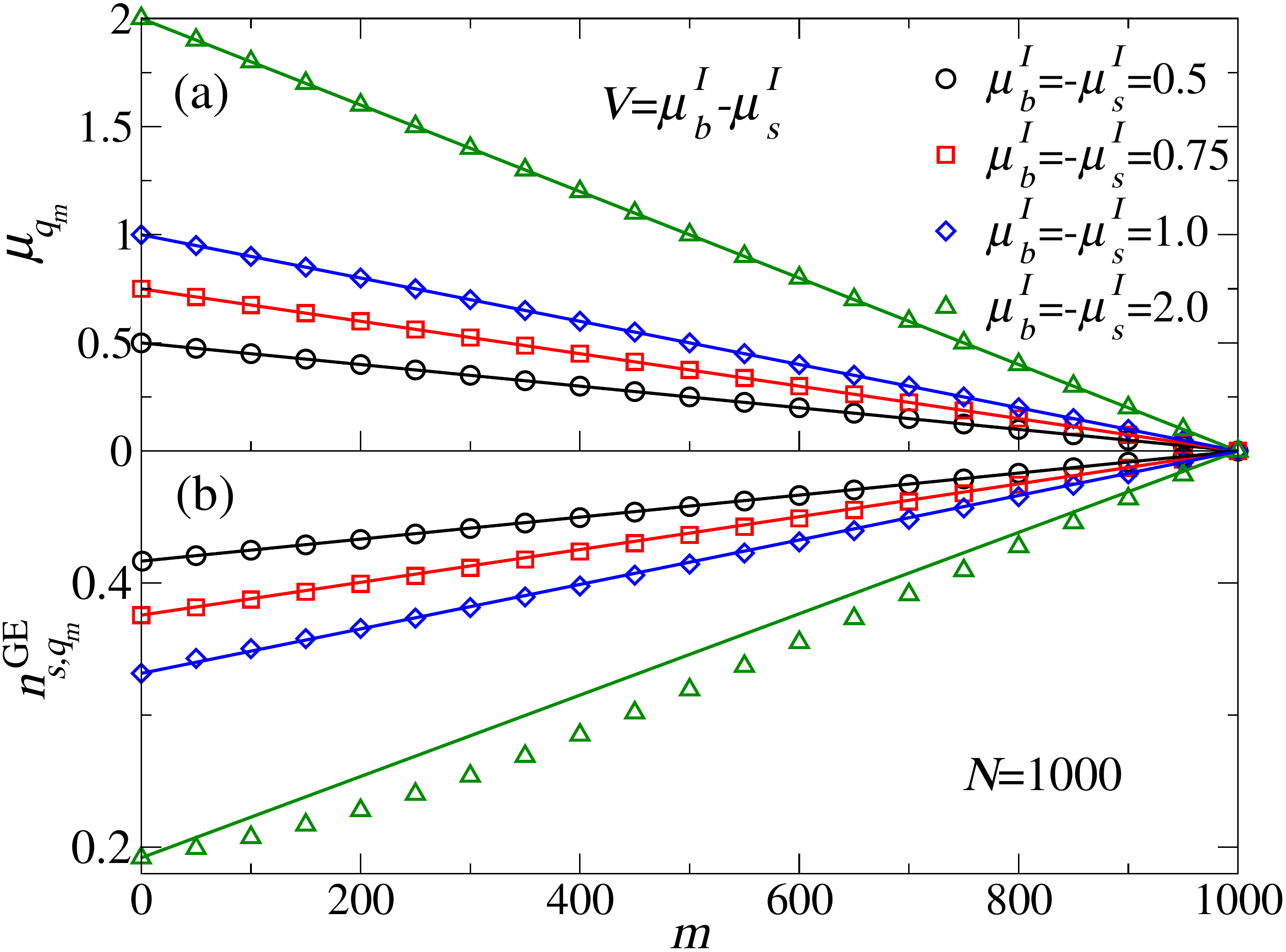}
\caption{(Color online) (a) Chemical potential of the entire ``thermalized'' system after the $m^\text{th}$ weak quench, $\mu_{q_m}$, vs $m$. (b) Average site occupation in the subsystem following thermalization after the $m^\text{th}$ weak quench, $n^\text{GE}_{s,q_m}$, vs $m$. Results are reported for $V=\mu^I_b-\mu^I_s=1.0$, 1.5, 2.0, and 4.0, where $\mu^I_b=-\mu^I_s$, $\beta^I=1.0$, and $L=1000$. The lines depict linear interpolations between the results after the strong quench ($m=0$) and the $N^\text{th}$ weak quench ($m=N$).}
\label{fig1}
\end{figure}

Since the on-site potential of the subsystem is reduced the same amount after each weak quench, the chemical potential in the GE after the $m^\text{th}$ weak quench ($\mu_{q_m}$) exhibits a linear decrease with $m$ (the temperature decreases slightly after every weak quench): $\mu_{q_m}\simeq \mu_{Q}+[(\mu_{F}-\mu_Q)m]/N$ [see Fig.~\ref{fig1}(a)], where $\mu_{Q}$ and $\mu_{F}$ are the chemical potentials after the strong quench and at the end of the cyclic process, respectively. Given $\mu_{q_m}$, the site occupations in the subsystem $n^\text{GE}_{s,q_m}$ can be obtained computing an integral such as the one in Eq.~\eqref{Eq.3}. The relation between $\mu_{q_m}$ and $n^\text{GE}_{s,q_m}$ is, in general, not a linear one. However, when $|\mu_{Q}|$ and $|\mu_{F}|$ are much smaller than the bandwidth, the relation is linear and 
\begin{equation}\label{eq:linearnmu}
 n^\text{GE}_{s,q_m}\simeq n^\text{GE}_{s,Q} +\frac{(n^\text{GE}_{s,F}-n^\text{GE}_{s,Q})m}{N},
\end{equation}
where $n^\text{GE}_{s,F}$ is the site occupation in the subsystem at the end of the cyclic process. This relation works remarkably well when $V\lesssim 2$ for $\mu^I_b=-\mu^I_s$ [see Fig.~\ref{fig1}(b)]. We choose $\mu^I_b=-\mu^I_s$ so that the system is at half-filling for $L_s=L_b$.

Using Eq.~\eqref{eq:linearnmu}, it is straightforward to compute the work extracted when ignoring the final local quench in which the subsystem and the bath are disconnected 
\begin{eqnarray}
&&W^\text{GE}(N)=E^I-E^\text{GE}_{F,N}\label{Eq:GEenergy} \\
&&\simeq VL_s\left(\frac{n^\text{GE}_{s,F}+n^\text{GE}_{s,Q}}{2}-n^I_s\right)-\frac{VL_s}{2N}\left(n^\text{GE}_{s,F}-n^\text{GE}_{s,Q}\right).\nonumber 
\end{eqnarray}

\begin{figure}[!t]
\includegraphics[width=0.48\textwidth]{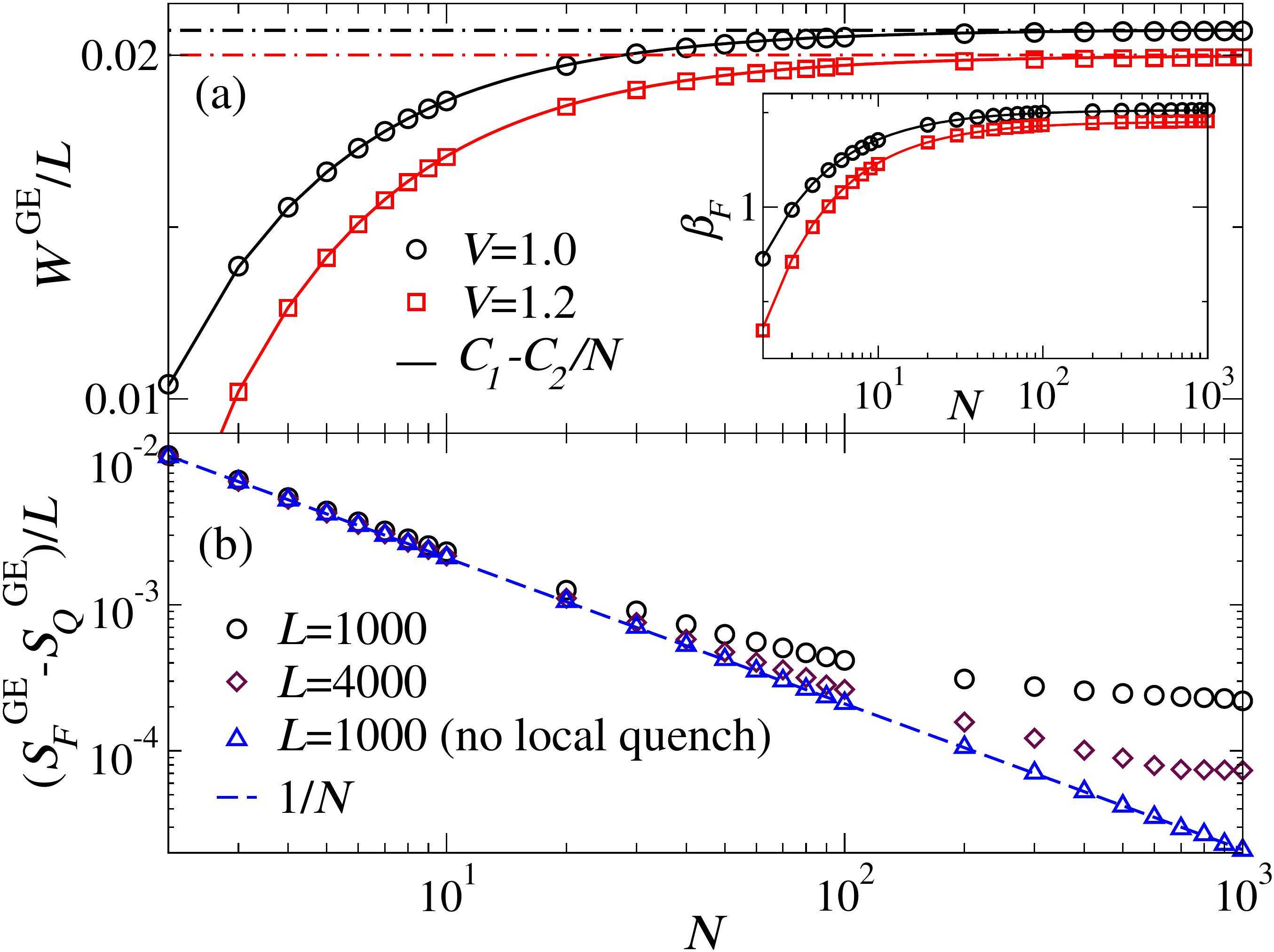}
\caption{(Color online) (a) Work extracted per site in the cyclic process in Fig.~\ref{fig0} when thermalization occurs, $W^\text{GE}/L$, vs the total number of small quenches $N$. Solid lines show the results from Eq.~\eqref{Eq:GEenergy}, while the dashed-dotted lines show the results from Eq.~\eqref{Eq:GEenergy} for $N\rightarrow\infty$. The results are for $L_s=L_b=500$ at half-filling, $\mu^I_s=-0.5$, $\mu^I_b=0.5$, $\beta^I=1$, and for quenches with $V=1.0$ (maximal work extraction) and $V=1.2$. (Inset) Inverse final temperature $\beta_F$ vs $N$. Solid lines correspond to a fit $C_1-C_2/N$. (b) $(S^\text{GE}_F-S^\text{GE}_Q)/L$ vs the total number of small quenches $N$ for $\mu^I_s=-0.5$, $\mu^I_b=0.5$, $\beta^I=1$, and for quenches with $V=1.0$. Results are presented when the final local quench is ignored (bottom curve with $1/N$ behavior highlighted by a dashed line,  $L_s=L_b=500$), and for two system sizes ($L_s=L_b=500$ and $L_s=L_b=2000$) in the cyclic process in Fig.~\ref{fig0}.}
\label{fig2}
\end{figure}

As shown in Fig.~\ref{fig2}(a), this expression is in excellent agreement with the exact numerical results for $L=1000$, $V=1.0$ and $1.2$, and $\mu^I_b=-\mu^I_s=0.5$, when the final local quench in which the subsystem and the bath are disconnected is taken into account. The top curve ($V=1.0$) shows results when the strong quench fulfills the condition for maximal work extraction. The vertical dashed-dotted lines depict the predictions of Eq.~\eqref{Eq:GEenergy} for $N\rightarrow\infty$.

The entropy in the GE can be written as~\cite{he2012initial}
\begin{equation}
S^\text{GE}=-\sum_{\alpha=1}^{L}[I^\text{GE}_\alpha \ln I^\text{GE}_\alpha +(1-I^\text{GE}_\alpha) \ln(1-I^\text{GE}_\alpha)]\, ,
\label{Eq.2}
\end{equation}
where $I_\alpha=(\exp[\beta(\epsilon_\alpha-\mu)]+1)^{-1}$ is the occupation of the single-particle states in the GE, and $\epsilon_\alpha$ are the single-particle eigenenergies. It follows from Eq.~\eqref{Eq.2} that the derivative of the GE entropy at the end of the cyclic process $S^\text{GE}_{F}$, with respect to the total number of weak quenches $N$, is $dS^\text{GE}_{F}/dN=\beta_F\sum_\alpha (\epsilon_\alpha-\mu_F)\,dI_\alpha/dN$, where the final inverse temperature $\beta_F$ and chemical potential $\mu_F$ depend on $N$. Since $\sum_\alpha I_\alpha$ equals the total number of particles, which is conserved during our cyclic process, $\sum_\alpha \mu_F\,dI_\alpha/dN=0$. Hence, as expected, $dS^\text{GE}_{F}/dN=\beta_F\sum_\alpha \epsilon_\alpha\,dI_\alpha/dN=\beta_F\, dE^\text{GE}_{F,N}/dN$. 

Using Eq.~\eqref{Eq:GEenergy}, we see that 
\begin{equation}\label{eq:entropyGE}
 \frac{dS^\text{GE}_{F}}{dN}=-\beta_F\frac{dW^\text{GE}(N)}{dN}\simeq 
 -\frac{VL_s\beta_F}{2N^2}\left(n^\text{GE}_{s,F}-n^\text{GE}_{s,Q}\right).
\end{equation}
For $N$ sufficiently large, $\beta_F$ and $n^\text{GE}_{s,F}$ are independent of $N$ [see the inset in Fig.~\ref{fig2}(a) for the behavior of $\beta_F$ vs $N$], and $S^\text{GE}_{F}\propto 1/N$.  

In Fig.~\ref{fig2}(b), we show the difference between the GE entropy per site at the end of the cyclic process $S_F^\text{GE}/L$ and the GE entropy per site after the strong quench $S_Q^\text{GE}/L$. Figure~\ref{fig2}(b) shows that, when the final local quench is ignored, the entropy difference vanishes with $1/N$ as predicted. When the last local quench is taken into account, the entropy difference can be seen to saturate with increasing $N$. As expected, with increasing system size, the effect of the local quench becomes negligible and the difference approaches the prediction in Eq.~\eqref{eq:entropyGE}. 

\subsection{Work Extraction and entropy differences vs the quench parameter}

\begin{figure}[!t]
\centering
\includegraphics[width=0.48\textwidth]{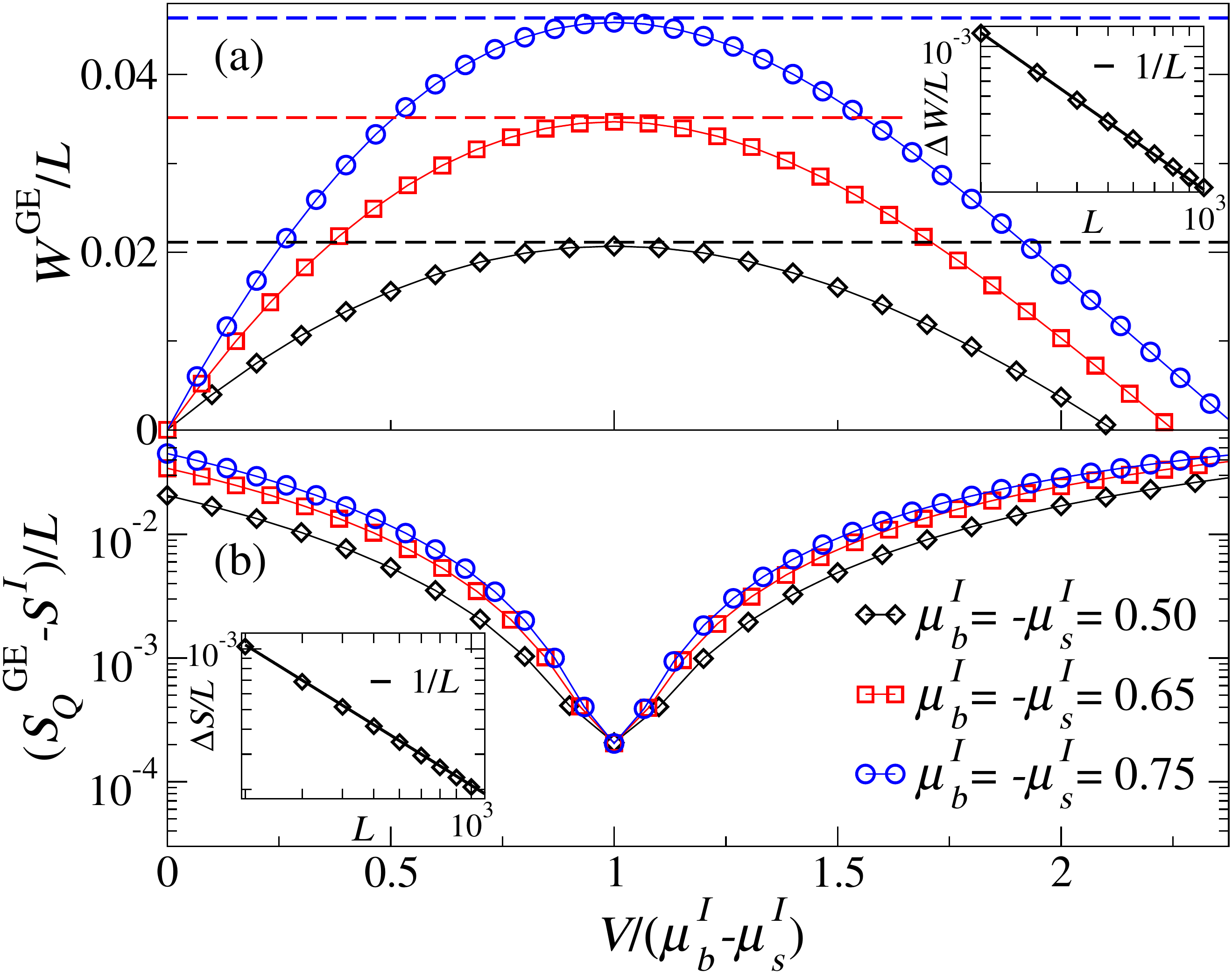}
\caption{(Color online) (a) Work extracted per site in the limit $N\rightarrow\infty$ in the cyclic process in Fig.~\ref{fig0}, $W^\text{GE}/L$ (see text), as a function of  $V/(\mu^I_b-\mu^I_s)$, for $L_s=L_b=500$, $\mu^I_b=-\mu^I_s=0.5$, $0.65$ and $0.75$, and $\beta^I=1.0$. The horizontal dashed lines show the maximal work bounds predicted by Eq.~\eqref{eq:maxwork}. (Inset) Difference between the maximal work bound and the numerical result for $W^\text{GE}$ at $V=\mu_b^I-\mu_s^I$, $\Delta W/L=|W^\text{GE}_\text{max}-W^\text{GE}|/L$, vs $L$ for $\mu^I_b=-\mu^I_s=0.5$. The solid line depicts a power law fit $\Delta W/L=a/L$ with $a=0.23$. (b) Difference between the GE entropy after the strong quench and the initial entropy, $(S^\text{GE}_{Q}-S^I)/L$ vs $V/(\mu^I_b-\mu^I_s)$, for the same parameters as in (a). (Inset) $(S^\text{GE}_{Q}-S^I)/L$ vs $L$, at $V=\mu_b^I-\mu_s^I$ for $\mu^I_b=-\mu^I_s=0.5$. The solid line depicts a power law fit $(S^\text{GE}_{Q}-S^I)/L=b/L$ with $b=0.21$.}
\label{fig3}
\end{figure}

Here we study the effect of changing the strong quench strength (set by the value of $V$) in the work extracted in the limit $N\to\infty$, as well as on the GE entropy after the strong quench, for finite system sizes. The energy at the end of the cyclic process for $N\to\infty$ is computed as follows: (i) We determine the entropy of the GE that has the same energy and number of particles as our system after the strong quench. (ii) We determine the GE that has the same entropy and number of particles determined in (i) but for the Hamiltonian $\hat{\mathcal H}(0,0)$ [see Eq.~\eqref{eq:finHam}], i.e., the Hamiltonian of the system after the $N\to\infty$ weak quenches. (iii) We compute the energy of the GE in (ii) after the local quench in which the subsystem and the bath are disconnected. The difference between the initial energy and the energy determined in (iii) is the work extracted in the limit $N\to\infty$.

In Fig.~\ref{fig3}(a), we plot results for the work extracted per site vs $V/(\mu_b^I-\mu_s^I)$ for three values of $\mu_b^I=-\mu_s^I$, for $L=1000$. The dashed lines are the maximal work bounds predicted by Eq.~\eqref{eq:maxwork}. The numerical results show that, as advanced for $V=\mu_b^I-\mu_s^I$, the work extracted for each value of $\mu_b^I=-\mu_s^I$ nearly saturates the maximal work bound. The inset in Fig.~\ref{fig3}(a) shows that the small difference between the numerical result and the bound, for $\mu_b^I=-\mu_s^I=0.5$, vanishes as $1/L$ with increasing system size. Figure~\ref{fig3}(a) also shows that increasing the difference between $\mu_b^I$ and $\mu_s^I$ increases the maximal work one can extract when the system thermalizes.

Figure~\ref{fig3}(b) shows the difference between the GE entropy after the strong quench and the initial entropy, per site, vs $V/(\mu_b^I-\mu_s^I)$. Results are shown for the same three values of $\mu_b^I=-\mu_s^I$ and system size as in Fig.~\ref{fig3}(a). The entropy difference per site can be seen to be minimal when $V=\mu_b^I-\mu_s^I$, and the inset shows that it vanishes as $1/L$ with increasing system size.
 
\section{Generalized Gibbs Ensemble\label{secIV}}

In this section, we study what happens when the systems are truly noninteracting. While this might appear to be a theoretical exercise of no relevance to experimental systems (or microscopic quantum devices), one should bear in mind that if interactions are very weak there exists the possibility that the dynamics for experimentally relevant time scales (operation times) is well described by a noninteracting Hamiltonian. The same can be said about systems that are interacting but close to some integrable point \cite{cazalilla_11}. For experimentally relevant time scales, their dynamics can be described by an integrable Hamiltonian and they do not thermalize, even if at very long times (not accessible in experiments) one expects that integrability-breaking effects result in thermalization. Beautiful experiments with ultracold atoms in 1D geometries have shown such a lack of thermalization \cite{kinoshita_wenger_06, gring_kuhnert_12, langen_erne_15}, while others have demonstrated that thermalization does occur in (nearly) isolated quantum systems if they are not close to integrable regimes \cite{trotzky_chen_12, kaufman_tai_16, clos_porras_16}.

As mentioned before, the breakdown of thermalization in integrable systems is due to the existence of an extensive number of nontrivial conserved quantities. In the noninteracting system of interest here, the conserved quantities ${\hat{I}_j}$ are the occupations of the single-particle eigenstates of the Hamiltonian after the quench (they are conserved because the particles do not interact with each other). There are as many of those as lattice sites, i.e., there is an extensive number of them. In integrable systems in general, and in our noninteracting system in particular, observables after equilibration are expected to be described by the GGE~\cite{rigol2007relaxation} (see Refs.~\cite{vidmar_rigol_16, essler_fagotti_16, cazalilla_chung_16, caux_16} for recent reviews).

The GGE density matrix~\cite{rigol2007relaxation}, which is obtained maximizing the entropy under the constraints imposed by the conserved quantities (following Jaynes \cite{jaynes_57a}) and has been justified microscopically in terms of generalized eigenstate thermalization \cite{cassidy2011generalized, vidmar_rigol_16, caux_essler_13}, can be written as
\begin{equation}
 \hat{\rho}^\text{GGE}=\frac{1}{Z_\text{GGE}} e^{-\sum_\alpha\lambda_\alpha\hat{I}_\alpha},
\end{equation}
where $Z_\text{GGE}=\Tr[\exp(-\sum_\alpha\lambda _\alpha \hat{I}_\alpha)]$ is the partition function of the GGE. The Lagrange multipliers $\lambda_\alpha$ are determined by the condition $\Tr[\hat{\rho}^\text{GGE}\hat{I}_{\alpha}] = I^I_{\alpha} \equiv \Tr[\hat{\rho}^I\hat{I}_{\alpha}]$, in which $\hat{\rho}^I$ is the density matrix of the initial (nonstationary) state. In the fermionic system of interest here~\cite{rigol2007relaxation}
\begin{equation} \label{Ikgge}
\lambda_\alpha = \ln\left(\frac{1- I^I_\alpha}{I^I_\alpha} \right),
\end{equation}
and the GGE entropy is
\begin{equation}
S^\text{GGE}=-\sum_{\alpha=1}^{L}[I^I_\alpha \ln I^I_\alpha +(1-I^I_\alpha) \ln(1-I^I_\alpha)]\,.
\label{eq:sGGE}
\end{equation}

Unlike for systems that thermalize and hence can be described by the GE, there is no simple way to determine the occupation of the single-particle eigenstates of the Hamiltonian after a quench. In addition, contrary to the GE, those occupations are not a monotonic function of the single-particle eigenenergies. As a result, within the GGE, the entropy is not necessarily a monotonic function of the energy. Therefore, the analytical arguments used in the context of the GE are not valid within the GGE. 

In what follows, we report and discuss numerical results for the cyclic protocol in Fig.~\ref{fig0} when we replace the exact density matrix of the system after equilibration by the GGE density matrix. As shown in Fig.~\ref{fig4}, within our protocol, numerical results for work extraction using exact dynamics (waiting random times after equilibration) and the GGE density matrix are in excellent agreement.
 
\subsection{Work Extraction\label{secIVa}}

Given our initial state, which is a product of GE density matrices, the first question we address is the effect that the number of weak quenches $N$ has on the work extracted in the cyclic process. As mentioned before, within the GGE, the entropy is not necessarily a monotonic function of the energy. This means that there is no \textit{a priori} reason to expect that a quasi-static return to the initial Hamiltonian, following the strong quench, allows us to extract the most work. Actually, for a specific non-passive initial state, in Ref.~\cite{perarnau2016work} a quasi-static process was shown not to be optimal for extracting work in a noninteracting fermionic system.

\begin{figure}[!b]
\includegraphics[width=0.48\textwidth]{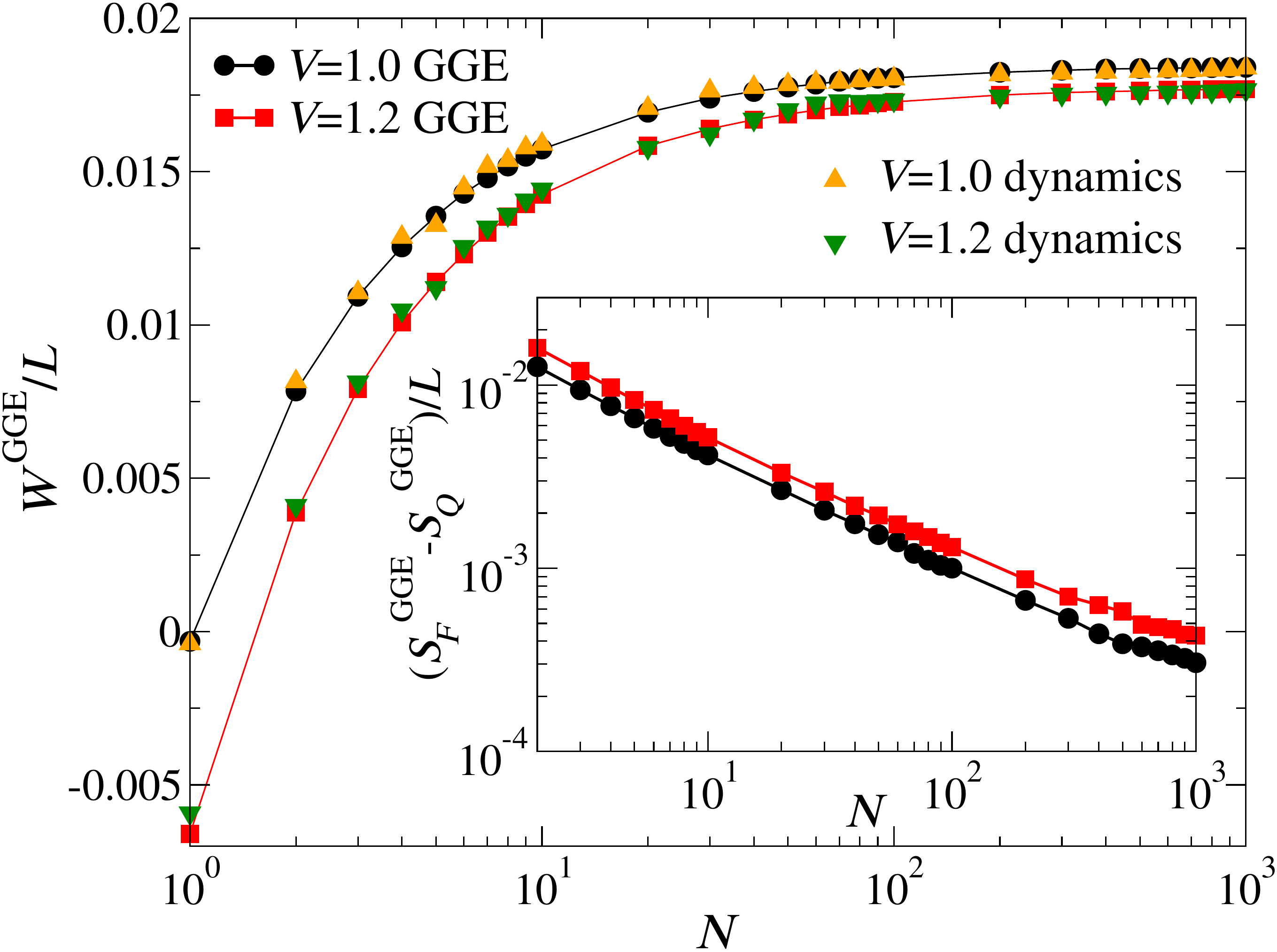}
\caption{(Color online) Work extracted per site within the exact dynamics (waiting random times after equilibration) and the GGE description, $W^\text{GGE}/L$, vs the total number of small quenches $N$. For the exact dynamics calculations, the random times selected are uniformly distributed between $200$ and $500$ (in units of inverse hopping) after each quench in 40 realizations of our cyclic process. (Inset) Difference between the GGE entropy per site at the end of the cyclic process and after the strong quench, $(S^\text{GGE}_{Q}-S^\text{GGE}_F)/L$, vs the total number of small quenches $N$. The final (local) quench is included in all calculations. The systems have $L_s=L_b=500$ ($L=1000$), are at half-filling ($\mu^I_b=-\mu^I_s=0.5$), and $\beta^I=1.0$. We report results for $V=1.0$ and $1.2$, as in Fig.~\ref{fig1}.}
\label{fig4}
\end{figure}

In Fig.~\ref{fig4}, we show the work extracted per site within the GGE description, $W^\text{GGE}/L$, for the same two cyclic processes as in Fig.~\ref{fig1}. Figure~\ref{fig4} shows that $W^\text{GGE}/L$ increases with $N$, and that, as $N\rightarrow\infty$, the work extracted for $V=1.0$ is greater than for $V=1.2$, as when thermalization occurs. The inset in Fig.~\ref{fig4} shows that the difference between the GGE entropy per site at the end of the cyclic process and after the strong quench, $(S^\text{GGE}_F-S^\text{GGE}_{Q})/L$, decreases with increasing $N$. Also, for any given $N$, the entropy difference is smaller for $V=1.0$ than for $V=1.2$. 

\begin{figure}[!b]
\includegraphics[width=0.48\textwidth]{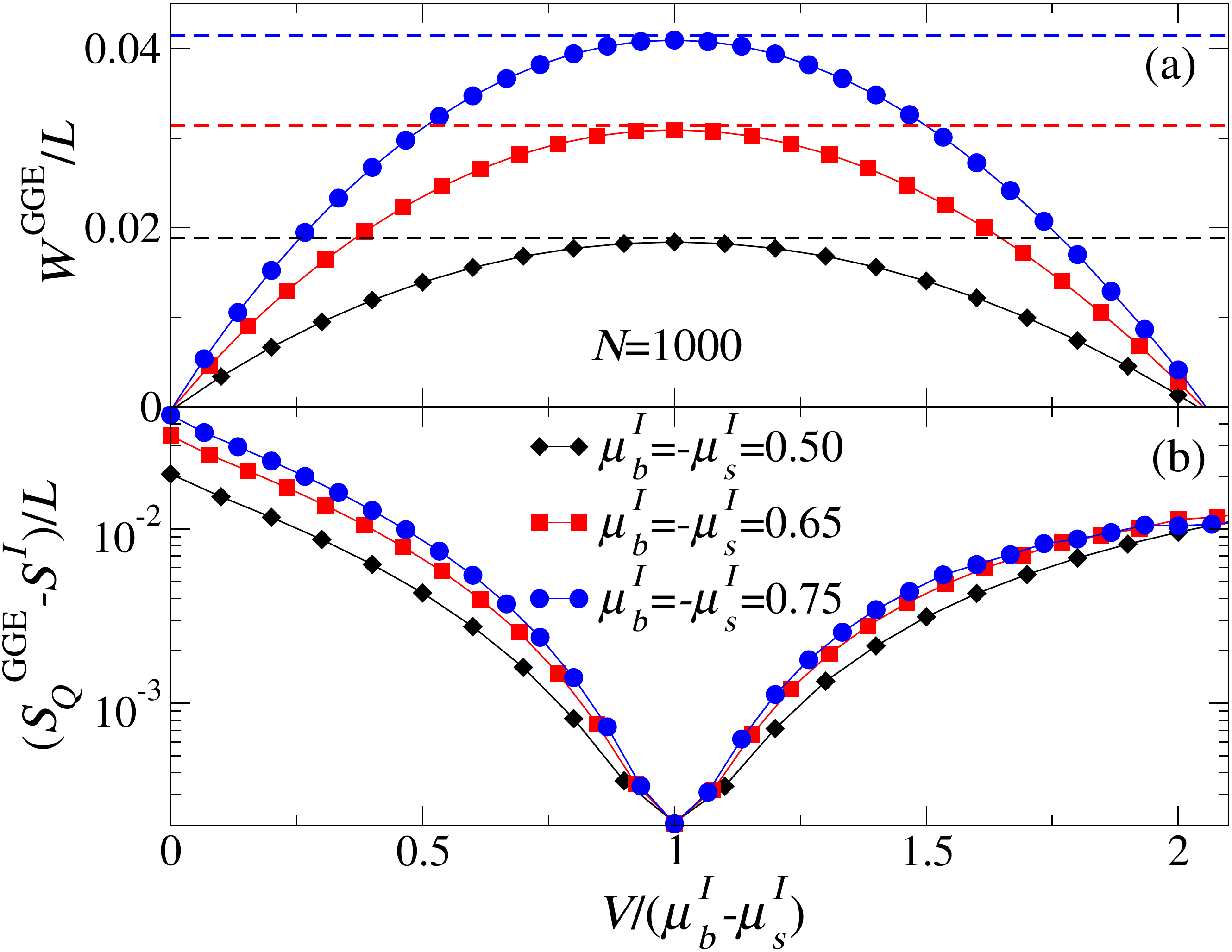}
\caption{(Color online) (a) Work extracted per site, $W^\text{GGE}/L$, in the cyclic process in Fig.~\ref{fig0} as a function of $V/(\mu^I_b-\mu^I_s)$, for $N=1000$, $L_s=L_b=500$, $\mu^I_b=-\mu^I_s=0.5$, $0.65$ and $0.75$, and $\beta^I=1.0$. The horizontal dashed lines show the maximal work bounds predicted by Eq.~\eqref{eq:maxworkGGE}. (b) Difference between the GGE entropy after the strong quench and the initial entropy, $(S^\text{GGE}_{Q}-S^I)/L$, vs $V/(\mu^I_b-\mu^I_s)$ for the same parameters as in (a).}
\label{fig5}
\end{figure}

Next, we study how changing the strength of the strong quench changes the work extracted within the GGE description for a large, but finite, number $N$ of weak quenches. We report results for $N=1000$. (Unlike for the GE, to determine the work extracted in the limit $N\to \infty$ within the GGE, one needs to do numerical calculations for finite $N$ and extrapolate the results to $N\to \infty$.) Figure~\ref{fig5} shows that, similarly to the results obtained for the GE, maximal work is extracted when $V=\mu^I_b-\mu^I_s$. This is understandable in terms of entropy production (or the lack thereof) as, for $V=\mu^I_b-\mu^I_s$, the strong quench in our protocol does not produce entropy within the GGE description. This follows from the inequalities $S^\text{GE}_Q\geq S^\text{GGE}_Q\geq S^I$. Since $S^\text{GE}_Q\simeq S^I$ for $V=\mu^I_b-\mu^I_s$, it then follows that $S^\text{GGE}_Q\simeq S^I$. It also follows that $S^\text{GE}_Q\simeq S^\text{GGE}_Q$, which, together with the fact that the energy of the GE and the GGE must match after the strong quench, hints that the GGE density matrix after the strong quench is very close to that of the GE (as both occupations of single-particle eigenstates and their ordering must match). Hence, the GGE density matrix after the strong quench is (almost) passive. 

Given a density matrix $\hat{\rho}^I$, the work that can be extracted within a GGE description of equilibration during a cyclic process has an upper bound~\cite{perarnau2016work} 
\begin{eqnarray}
 W^\text{GGE}_\text{max}=\Tr[\hat{\rho}^I\hat{\mathcal H}^I]-\sum _\alpha \epsilon_\alpha I'_\alpha,
\label{eq:maxworkGGE}
\end{eqnarray}
where $\epsilon_\alpha$ are the single-particle energy eigenvalues of $\hat{\mathcal H}^I$ (in ascending order), and $I'_\alpha$ are the occupations of single-particle eigenstates in $\hat{\rho}^I$ reordered in descending order so that $\sum_\alpha \epsilon_\alpha I'_\alpha$ is the minimal energy given that set of occupations. Since the occupations of single-particle energy eigenstates are the same in the initial state and in the hypothetical final state with minimal energy, both states have identical entropies. 

Figure~\ref{fig5} shows that the maximal work extracted in our cyclic process is very close to that bound. As mentioned before, for $V=\mu^I_b-\mu^I_s$ and $N\rightarrow\infty$, our protocol ensures that $S_F^\text{GGE}\simeq S^I$, i.e., the initial and final sets of occupations of the single-particle eigenstates of $\hat{\mathcal H}^I$ are expected to be the same. In addition, since the density matrix of the GGE after the strong quench is (almost) passive, $N\rightarrow\infty$ ensures that the density matrix of the GGE at the end of the cyclic process is (almost) passive, i.e., our final state is (almost)  the hypothetical final state in Eq.~\eqref{eq:maxworkGGE}.

\begin{figure}[!t]
\includegraphics[width=0.48\textwidth]{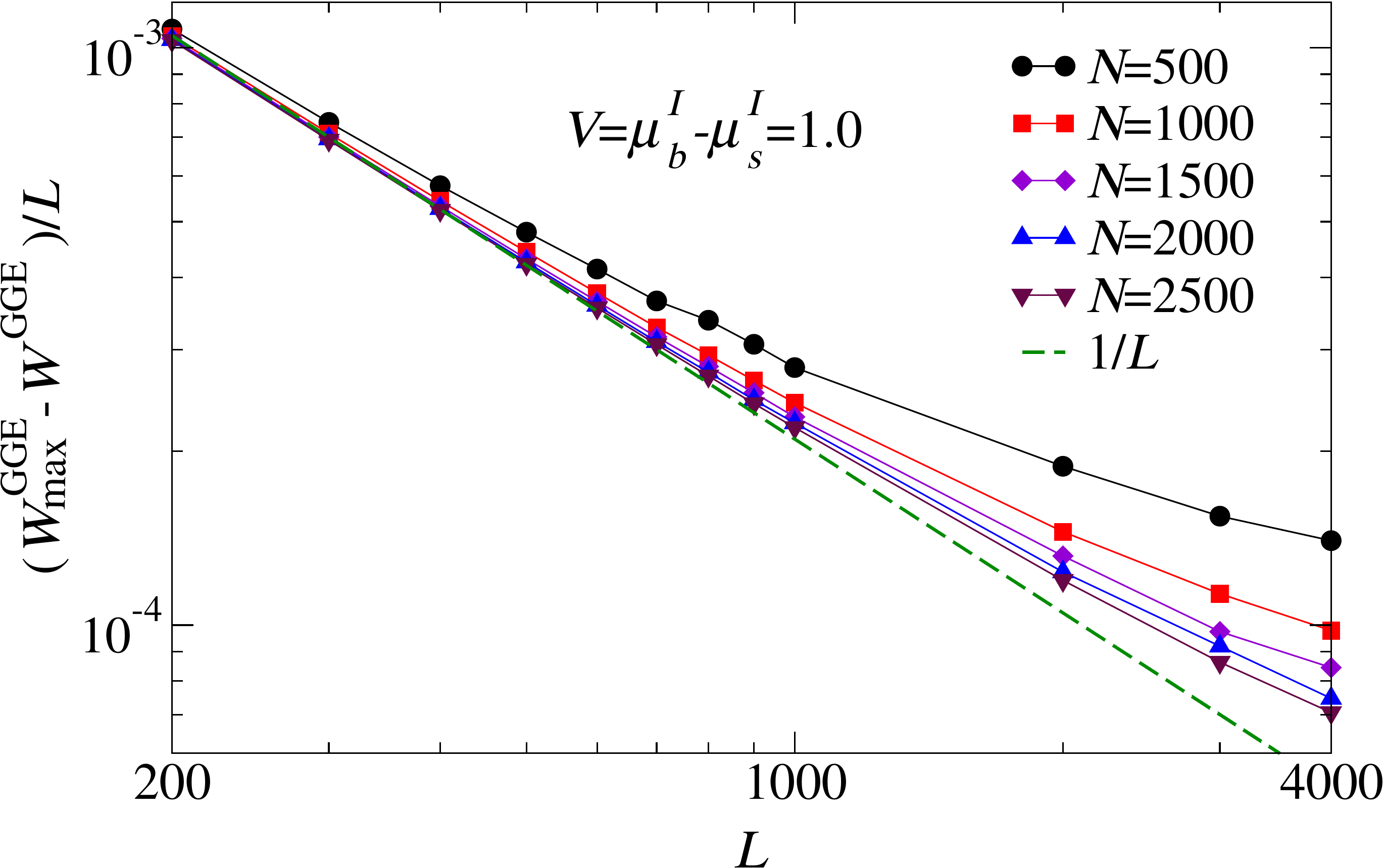}
\caption{(Color online) Difference $\Delta W^\text{GGE}/L =(W^\text{GGE}_\text{max}-W^\text{GGE})/L$, at $V=\mu^I_b-\mu^I_s$, vs $L$ ($L_s=L_b$). We present results for different number of weak quenches ($N=500$, $1000$, $1500$, $2000$, and $2500$), for systems at half filling ($\mu^I_b=-\mu^I_s=0.5$), and $\beta^I=1.0$. The dashed line depicts a $1/L$ scaling.} 
\label{fig6}
\end{figure}

We have studied what happens with the small differences seen in Fig.~\ref{fig5} between the numerical calculations using the GGE and the predictions of Eq.~\eqref{eq:maxworkGGE}, $\Delta W^\text{GGE}/L=(W^\text{GGE}_\text{max}-W^\text{GGE})/L$ at $V=\mu^I_b-\mu^I_s$, when one changes the system size $L$ and the total number of small quenches $N$. In Fig.~\ref{fig6}, we show results for $V=\mu^I_b-\mu^I_s=1.0$. As expected, $\Delta W^\text{GGE}/L$ decreases with increasing $N$ and $L$. With increasing $N$, the results approach a power law decay $\propto 1/L$. This confirms our expectation that as $N\rightarrow\infty$ and then $L\rightarrow\infty$, our protocol for $V=\mu^I_b-\mu^I_s$ saturates the bound in Eq.~\eqref{eq:maxworkGGE}.

\section{Comparison between GE and GGE descriptions\label{secV}}

In this section, we discuss the differences between the GE and the GGE descriptions of our cyclic process. We should stress that, after a quench starting from the same initial state, the energy and number of particles within the GE and GGE descriptions are identical. What are different are the density matrices describing the system. This is what leads to different results after subsequent quenches and, ultimately, to different work extracted after completing cyclic processes.

As we discussed in Secs.~\ref{secIII} and~\ref{secIV}, the protocols devised to extract maximal work within the GE and GGE descriptions do not increase the entropy of the system. Since in the GE the energy is a monotonically increasing function of the entropy, and for any given energy the entropy is maximal, whenever the GGE has the same entropy as the GE it must have a higher energy. As a result, $W^\text{GE}_\text{max} \geq W^\text{GGE}_\text{max}$. This can be seen if one compares the results in Fig.~\ref{fig3}(a) and in Fig.~\ref{fig5}(a). A recent work has proposed a protocol to extract $W^\text{GE}_\text{max}-W^\text{GGE}_\text{max}$ in a noninteracting setting~\cite{verstraelen2017unitary}. 

\begin{figure}[!t]
\includegraphics[width=0.48\textwidth]{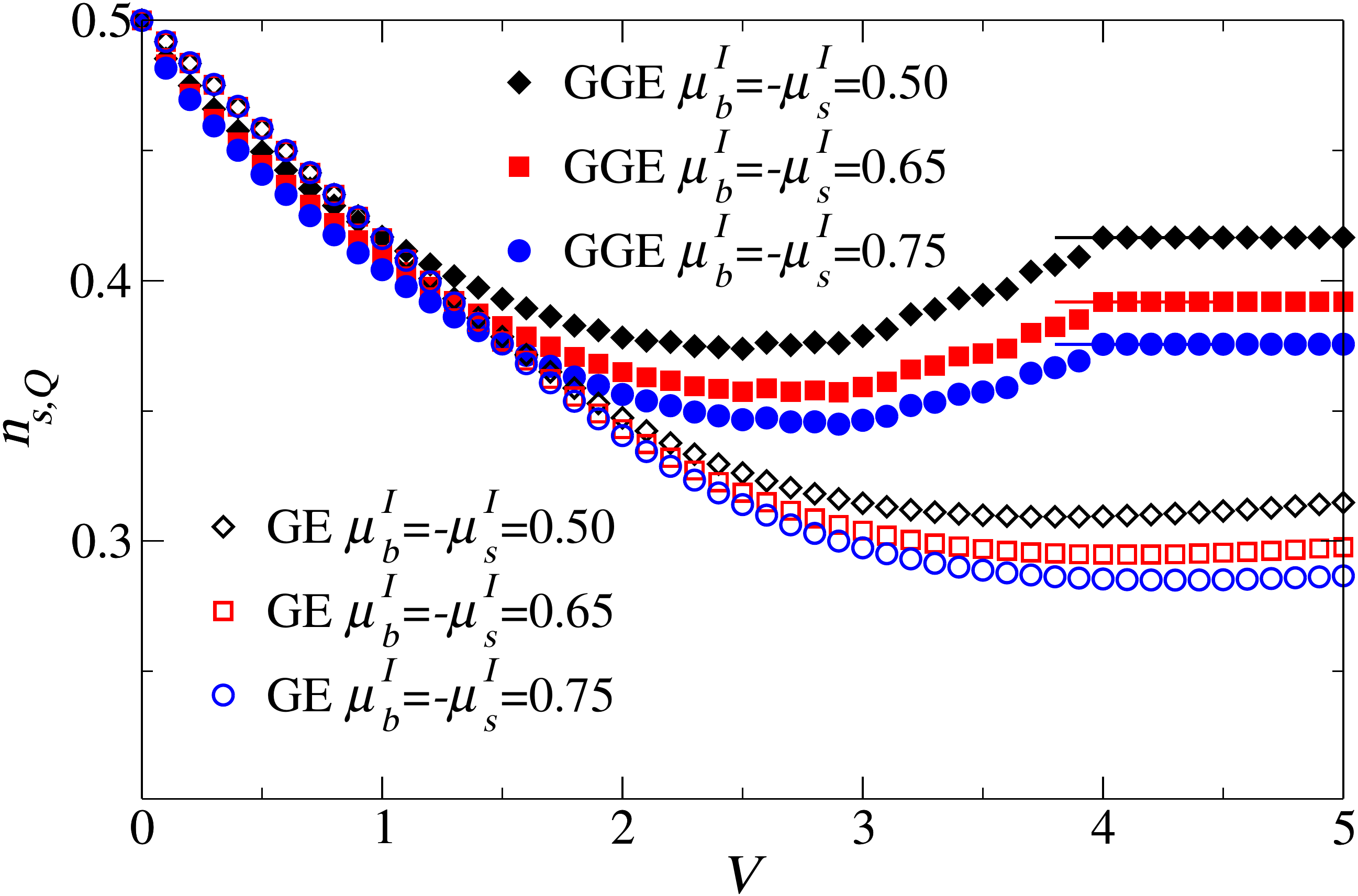}
\caption{(Color online) Average site occupation in the subsystem after the strong quench as a function of $V$ for equilibration to the GE and GGE descriptions. Results are shown for $\mu^I_b=-\mu^I_s$=$0.5$, $0.65$ and $0.75$, $L_s=L_b=500$, and $\beta^I=1.0$. The horizontal lines following the GGE results for $V\geq4$ make apparent that the site occupations in the subsystem are independent of $V$ in that regime.}
\label{fig7}
\end{figure}

A quantity that exhibits a qualitatively different behavior in the GE and GGE with increasing the quench strength is the average site occupation in the subsystem $n_{s,Q}$ (and, consequently, in the bath). In Fig.~\ref{fig7}, we plot $n_{s,Q}$ as a function of $V$ for three values of $\mu^I_b=-\mu^I_s$ (i.e., at half-filling). While one can see that in the GE, for $V\leq 4$, $n_{s,Q}$ decreases smoothly as $V$ increases, $n_{s,Q}$ exhibits a nonmonotonic behavior in the GGE with a sort of kink at $V=4$. For $V\geq4$, $n_{s,Q}$ does not change in the GGE with increasing $V$. This is because for $V\geq4$ all fermions that are in the subsystem (bath) before the quench remain in the subsystem (bath) after the quench, which is the result of the subsystem and the bath having a local on-site potential difference that is larger than the band-width, i.e., because of energy conservation the fermions cannot hop between the subsystem to the bath in the absence of interactions. 

\begin{figure}[!t]
\includegraphics[width=0.48\textwidth]{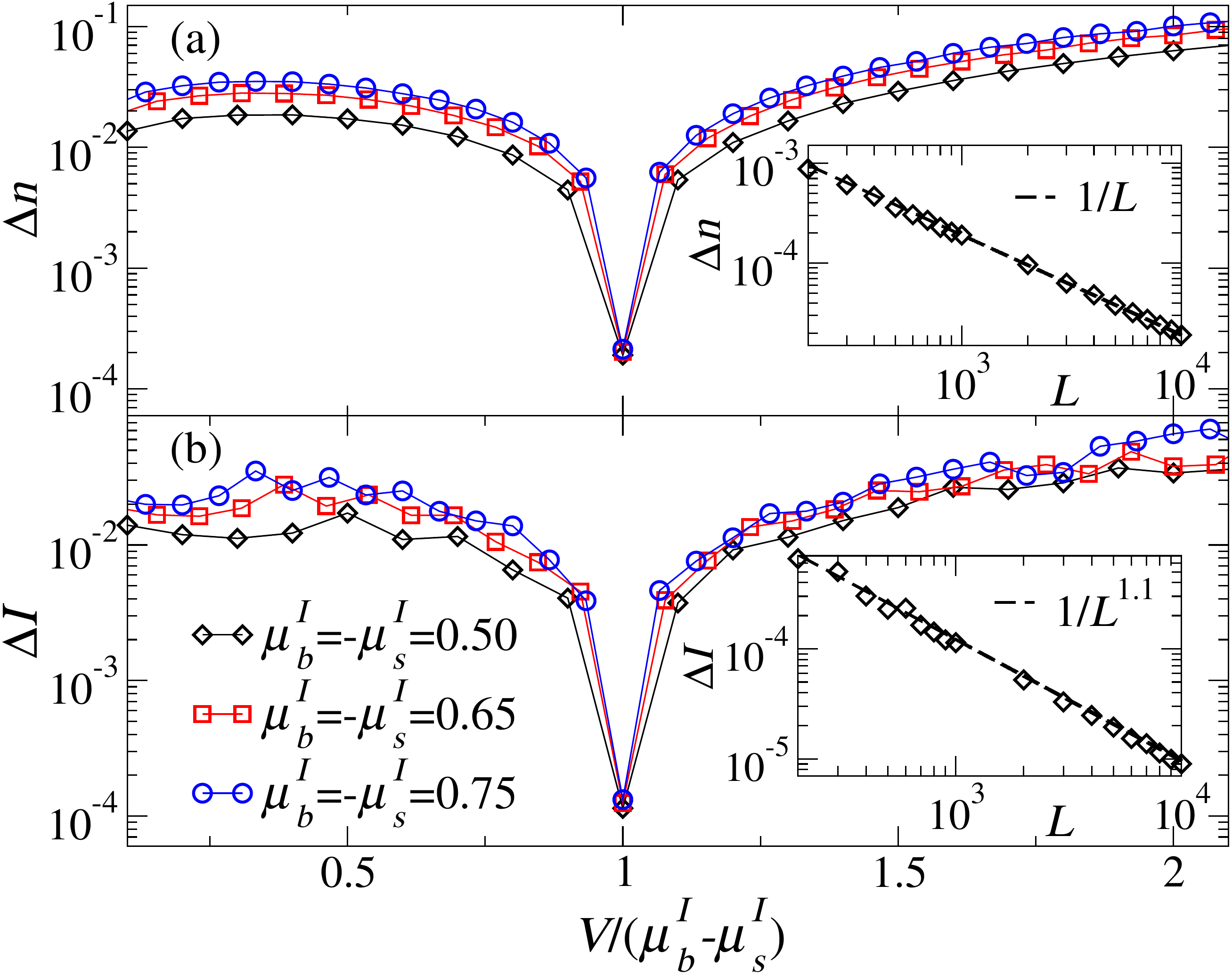}
\caption{(Color online) (a) Relative difference between the GE and GGE predictions for the site occupations in the system after equilibration following the strong quench, $\Delta n$ [see Eq.~\eqref{eq:reldiff}], vs $V/(\mu^I_b-\mu^I_s)$ for $\mu^I_b=-\mu^I_s=0.5$, $0.65$, and $0.75$, $L_s=L_b=500$, and $\beta^I=1.0$. (Inset) $\Delta n$ vs $L$ for $V=\mu^I_b-\mu^I_s=1.0$ and $\beta^I=1.0$. The dashed line depicts a power-law decay $\propto1/L$. (b) Relative difference between the occupation of the single-particle energy eigenstates in the GE and GGE following the strong quench, $\Delta I$ [see Eq.~\eqref{eq:reldiff}], vs $V/(\mu^I_b-\mu^I_s)$ for the same parameters as in (a). (Inset) $\Delta I$ vs $L$ for $V=\mu^I_b-\mu^I_s=1.0$ and $\beta^I=1.0$. The dashed line depicts a power-law decay $\propto1/L^{1.1}$.}
\label{fig8}
\end{figure}

In order to be more quantitative in the comparison of the GE and the GGE, we calculate the relative differences in the site ($n^\text{GE/GGE}_{i,Q}$) and single-particle energy eigenstate ($I^\text{GE/GGE}_{\alpha,Q}$) occupations after the strong quench:
\begin{eqnarray}
 \Delta n=\frac{\sum_i |n^\text{GE}_{i,Q}-n^\text{GGE}_{i,Q}|}{\sum_i n^\text{GGE}_{i,Q}}, \nonumber \\
 \Delta I=\frac{\sum_\alpha |I^\text{GE}_{\alpha,Q}-I^\text{GGE}_{\alpha,Q}|}{\sum_\alpha I^\text{GGE}_{\alpha,Q}}, \label{eq:reldiff}
 \end{eqnarray}
Results for those two quantities are reported in Fig.~\ref{fig8}. One can see there that $\Delta n$ [Fig.~\ref{fig8}(a)] and $\Delta I$ [Fig.~\ref{fig8}(b)] have deep minima (note the logarithmic scale) at $V=\mu^I_b-\mu^I_s$, which correspond to the quenches for which maximal work is extracted within both ensemble descriptions. As discussed in Sec.~\ref{secIVa}, for such strong quenches the GE and GGE density matrices are expected to be very close to each other. The insets in Fig.~\ref{fig8} show that, as expected, $\Delta n$ and $\Delta I$ for $V=\mu^I_b-\mu^I_s$ vanish with increasing system size (almost linearly with $1/L$).

\begin{figure}[!t]
\includegraphics[width=0.48\textwidth]{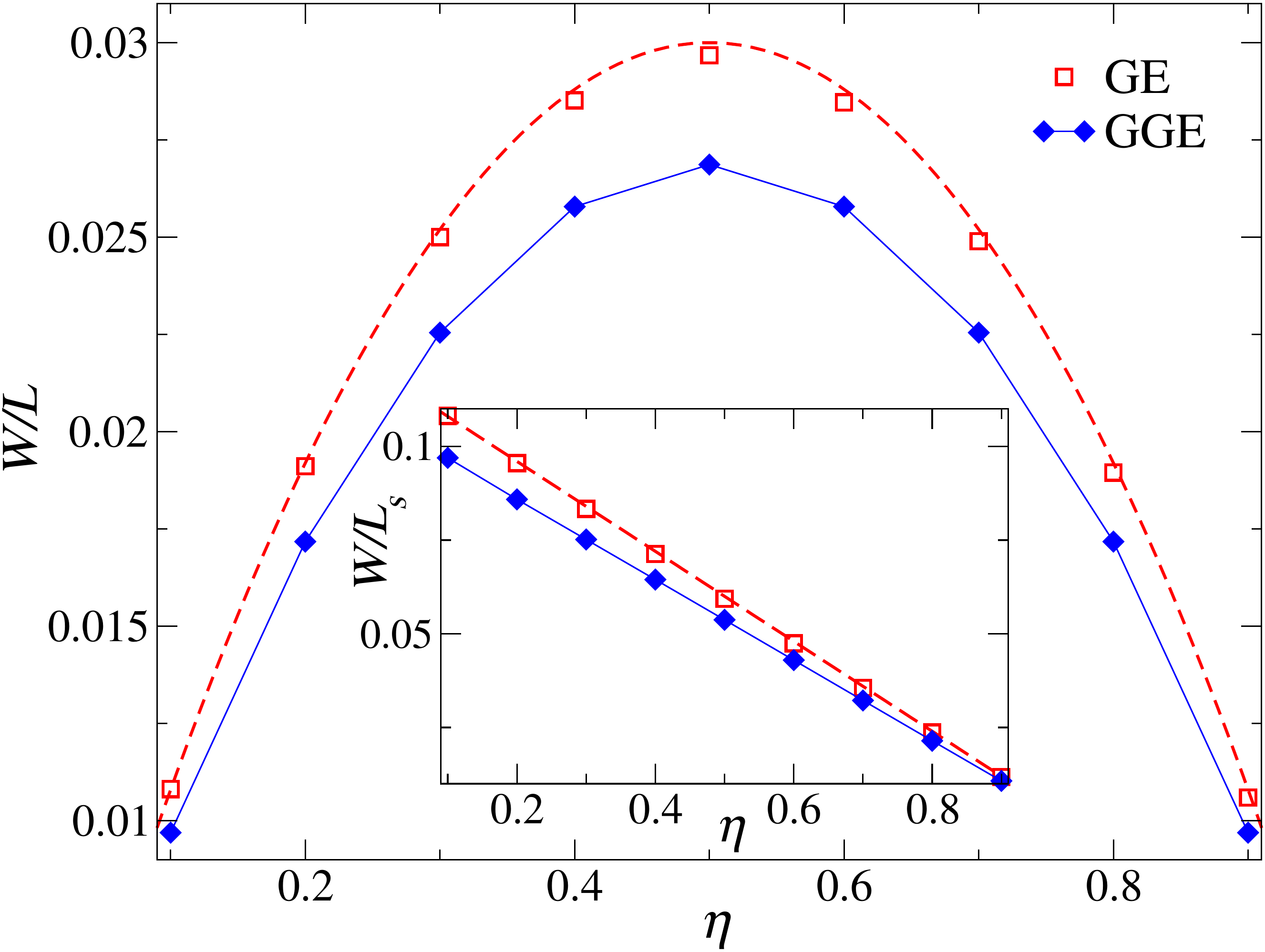}
\caption{(Color online) (Main panel) Work extracted divided by the system size, $W/L$, vs $\eta$, and (inset) work extracted divided by the subsystem size, $W/L_s$, vs $\eta$ for the descriptions within the GE ($N\to\infty$) and the GGE ($N=1000$). Results are shown for $L=1000$, $\beta^I=1.0$, and $V=\mu^I_b-\mu^I_s=1.2$, where $\mu^I_b=-\mu^I_s=0.6$. Dashed lines are the analytical prediction from Eq.~\eqref{w_eta}, while the continuous lines are a guide to the eye.}
\label{fig9}
\end{figure}

In all numerical results reported so far, we considered the case in which the subsystem and the bath have the same size, $L_s=L_b$. In Fig.~\ref{fig9}, we report results obtained when changing the ratio between the size of the subsystem and the size of the entire system, $\eta=L_s/L$. We focus on the protocol for which maximal work can be extracted, $V=\mu^I_b-\mu^I_s$. Figure~\ref{fig9} shows that, both for the GE and GGE descriptions, the work extracted per site in the entire system is maximal when $L_s=L_b$ ($\eta=1/2$). On the other hand, the inset in Fig.~\ref{fig9} shows that the work extracted per site in the subsystem is a monotonically decreasing function of $\eta$. Again, this is true both for the GE and GGE. Depending on whether one wants to extract the most work or the most work per site of the subsystem, one needs to select the subsystem size to be equal to that of the bath or much smaller than that of the bath, respectively.

The GE results in Fig.~\ref{fig9} can be understood in the context of the theoretical framework discussed in Sec.~\ref{secIII} for the thermodynamic limit. The strong quench with $V=\mu^I_b-\mu^I_s$ does not produce entropy, and is used to extract maximal work. For this quench, the chemical potential and temperature of the entire system in the GE that describes the thermalized observables are $\mu_{Q}=\mu^I_b$ and $\beta_{Q}=\beta^I$, respectively. The site occupations in the subsystem remain unchanged from their initial values. Hence, substituting $n^\text{GE}_{s,Q}=n^I_s$, $V=\mu^I_b-\mu^I_s$, and $N \to \infty$ in Eq.~\eqref{Eq:GEenergy}, one obtains $W^\text{GE}\simeq L_s (\mu^I_b-\mu^I_s)(n^{s}_{F}-n^I_s)/2$. The average site occupation in the final state is $n^\text{GE}_{s,F}=\eta\, n^I_s +(1-\eta)n^I_b$, where $\eta=L_s/L$, so the (maximal) work extracted in our cyclic process is
\begin{eqnarray}
\frac{W^\text{GE}}{L}  &\simeq& (\mu^I_b-\mu^I_s) (n^I_b-n^I_s)\eta (1-\eta)/2, \nonumber \\
\frac{W^\text{GE}}{L_s}&\simeq& (\mu^I_b-\mu^I_s) (n^I_b-n^I_s)     (1-\eta)/2. 
\label{w_eta}
\end{eqnarray}
The dashed lines in the main panel in Fig.~\ref{fig9} and its inset depict the results from Eq.~\eqref{w_eta}, and can be seen to be in excellent agreement with the numerical results for $N\rightarrow\infty$ and $L=1000$.

\begin{figure}[!t]
\includegraphics[width=0.48\textwidth]{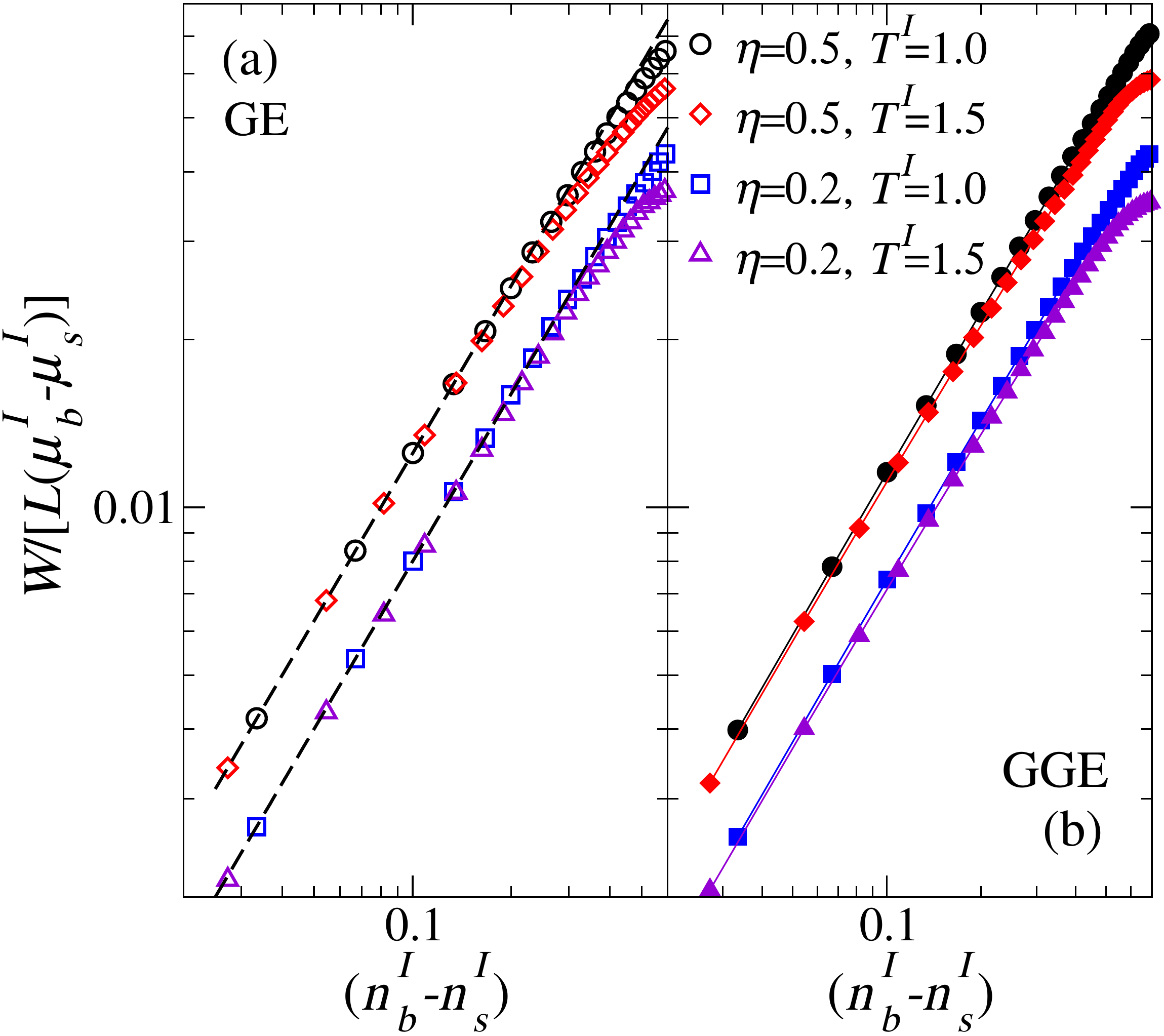}
\caption{ (Color online) $W/[L(\mu^I_b-\mu^I_s)]$ vs $(n^I_b-n^I_s)$ within (a) the GE ($N\to \infty$) and (b) the GGE ($N=1000$). Results are reported for $L=1000$, for two initial temperatures ($T^I=1.0$ and $1.5$), and two subsystem sizes ($\eta=0.5$ and $0.2$). The dashed lines in (a) are results from Eq.~\eqref{w_eta}, while the continuous lines in (b) are a guide to the eye. All results were obtained for $V=\mu^I_b-\mu^I_s$ and $\mu^I_b=-\mu^I_s$.}
\label{fig10}
\end{figure}

In Fig.~\ref{fig10}, we plot $W/[L(\mu^I_b-\mu^I_s)]$ vs $n^I_b-n^I_s$ for two values of $\eta$ and for two initial temperatures $T^I=(\beta^I)^{-1}$. Results within the GE description are shown in Fig.~\ref{fig10}(a), while results within the GGE description are shown in Fig.~\ref{fig10}(b). The results in Fig.~\ref{fig10}(a) are in excellent agreement with the predictions of Eq.~\eqref{w_eta} whenever the difference in site occupations in the subsystem and the bath, as well as the initial temperature, are not too large. They allow one to also gain a qualitative understanding of what happens within the GGE description because, as seen in Fig.~\ref{fig10}(b), the GGE results are qualitatively similar to those obtained within the GE. We note that, both in the GE and GGE, $W/[L(\mu^I_b-\mu^I_s)]$ for a given value of $n^I_b-n^I_s$ decreases with increasing $T^I$.

\section{Summary\label{secVI}}

We studied work extraction within a fermionic quadratic model in an isolated (unitarily evolving) 1D lattice system, and considered equilibration both to the GE, to describe what happens in a weakly interacting quantum chaotic system, as well as to the GGE, to describe what happens in the noninteracting limit. We considered initial states that are products of thermal states of the subsystem and the bath (in which we divided the isolated system). We devised a cyclic protocol that begins by connecting the subsystem and the bath, and quenching the local on-site potentials in the subsystem (we called that quench the ``strong'' quench in our protocol). After equilibration, we applied a quasi-static process in which the local on-site potentials in the subsystem are brought to the values before the strong quench by means of $N$ weak quenches, after which the subsystem and the bath are disconnected (a local quench).

We calculated the work extracted when changing the strength $V$ of the strong quench, and the number $N$ of weak quenches, both within the GE and the GGE. We found the value of $V$ for which maximal work can be extracted both in the GE and the GGE, and discussed why for that value of $V$ our cyclic protocols saturate the theoretical bounds. We studied the effect of changing the ratio between the sizes of the subsystem and the bath, but focused in the case in which they are the same. This might be of interest to understand microscopic devises, and is different from the subsystem-much-larger-than-bath approach in traditional thermodynamics. Within our cyclic protocol, and in the parameter regime studied, no qualitative differences were found in the work extracted when considering equilibration to the GE and the GGE. Exploration of initial states for which systems described by the GGE can be used to extract work without involving quasi-static processes~\cite{perarnau2016work} is a topic of future research.

\begin{acknowledgements}
This work was supported by the Army Research Office Grant No. W911NF1410540. The computations were done at the Institute for CyberScience at Penn State.
\end{acknowledgements}
\bibliographystyle{biblev1}
\bibliography{reference}
\end{document}